\documentclass[a4paper,fleqn,usenatbib]{mnras}

\usepackage{newtxtext,newtxmath}

\usepackage[T1]{fontenc}
\usepackage{ae,aecompl}

\usepackage{graphicx}	
\usepackage{amsmath}	
\usepackage{amssymb}	
\usepackage{enumitem}

\title[Simulating coupled dust with {\sc Phantom}]{Enforcing dust mass conservation in 3D simulations of tightly-coupled grains with the {\sc Phantom} SPH code}

\author[Ballabio et al.]{\parbox{\textwidth}{
G.~Ballabio,$^{1}$\thanks{E-mail: gb258@leicester.ac.uk}
G.~Dipierro$^1$, 
B.~Veronesi$^2$, 
G.~Lodato$^2$, 
M.~Hutchison$^{3,4}$,
G.~Laibe$^5$ 
and D.~J.~Price$^6$}\vspace{0.2cm}\\ 
$^{1}$Department of Physics and Astronomy, University of Leicester, Leicester, LE1 7RH, UK\\
$^{2}$Dipartimento di Fisica, Universit\`a degli Studi di Milano, Milano, Italy\\
$^{3}$Physikalisches Institut, Universit{\"a}t Bern, Gesellschaftstrasse 6, 3012 Bern, Switzerland \\
$^{4}$Institute for Computational Science, University of Zurich, Winterthurerstrasse 190, CH-8057 Z{\"u}rich, Switzerland \\
$^{5}$Univ Lyon, Ens de Lyon, CNRS, Centre de Recherche Astrophysique de Lyon, Saint-Genis-Laval, France\\
$^{6}$Monash Centre for Astrophysics (MoCA) and School of Physics and Astronomy, Monash University, Australia
}

\date{Accepted XXX. Received YYY; in original form ZZZ}

\pubyear{2018}

\begin{document}
\label{firstpage}
\pagerange{\pageref{firstpage}--\pageref{lastpage}}
\maketitle

\begin{abstract}
We describe a new implementation of the one-fluid method in the SPH code {\sc Phantom} to simulate the dynamics of dust grains in gas protoplanetary discs. We revise and extend previously developed algorithms by computing the evolution of a new fluid quantity that produces a more accurate and numerically controlled evolution of the dust dynamics. Moreover, by limiting the stopping time of uncoupled grains that violate the assumptions of the terminal velocity approximation, we avoid fatal numerical errors in mass conservation. We test and validate our new algorithm by running 3D SPH simulations of a large range of disc models with tightly- and marginally-coupled grains.
\end{abstract}




\section{Introduction}
Protoplanetary discs are composed of a mixture of gas and dust. While gas usually dominates the mass, and hence the hydrodynamics of the system, dust is the dominant source of opacity in the bulk of the disc. As a result, the optical appearance of discs is strongly influenced by the dust distribution \citep{testi14a,2016Birnstiel}. Recent high-resolution observations of protoplanetary discs have revealed a wealth of asymmetric structures in both gas and dust phases \citep[e.g.][]{2016PASA...33...13C,2018ApJ...853..162B}. The physical mechanisms driving the formation of these structures are best understood using 3D hydrodynamical simulations that accurately model the coupling between gas and dust for a wide range of grain sizes \citep{haworth16a}.

Solid particles embedded in a gas fluid are often treated using a continuous fluid description \citep{garaud04}. 
The macroscopic properties of the dust (e.g. density and velocity) are evolved on a set of grid points or particles that represent a volume large enough to be statistically meaningful, but sufficiently small as to ignore variations of the fluid quantities within that volume.
In Smoothed Particle Hydrodynamic (SPH) simulations, the dust dynamics can be computed using two different approaches: the two-fluid algorithm described in \citet{2012MNRAS.420.2345L}, typically used for large dust grains in weak drag regimes, and the one-fluid algorithm \citep{2015MNRAS.451..813P} based on the so-called terminal velocity approximation \citep{2005ApJ...620..459Y}, which is better suited for simulating dust phases that are tightly coupled with the gas. In terms of gas-dust modelling, the two-fluid implementation treats the gas and the dust as two separate sets of simulation particles, coupled by a drag force. In contrast, the SPH particles in the one-fluid approach represent the mixture, whose composition is determined by the dust fraction, that is evolved as a local property of the mixture.

Since it is numerically difficult to simulate all of the physical drag regimes that occur in nature with a single algorithm, methods/studies are often distinguished by the degree of coupling between phases, usually quantified by the so-called \emph{Stokes number}: particles with the same Stokes number are aerodynamically identical -- regardless of their shape, size, and/or density. The Stokes number is found by comparing the typical dynamical timescale of the system, $t_{\rm dyn}$, to the typical stopping timescale, $t_{\rm s}$, i.e. the time it takes for drag to significantly modify the relative velocity between a single grain and the gas. When the grains size is smaller than the mean free path of the gas \citep{epstein24a} -- which is generally the case for mm-cm size grains in protoplanetary discs \citep{garaud04} --, the Stokes number is given by \citep{2015MNRAS.451..813P}
\begin{equation}
\mathrm{St} \equiv \frac{t_{\rm s}}{t_{\rm dyn}}= \sqrt{\frac{\pi \gamma}{8}}\frac{ \rho_{\rm int} a \Omega_{\mathrm{k}}}{\rho c_{\rm s}l} ,
\end{equation}
where $a$ is the grain size, $\rho_{\rm int}$ is the intrinsic grain density, $c_{\rm s}$ is the sound speed, $\gamma$ is the adiabatic index, $\Omega_{\mathrm{k}}$ is the Keplerian angular velocity ($\Omega_{\rm k} = t_{\rm dyn}^{-1}$), $l$ is a correction factor for supersonic drag and $\rho$ is the total density. 


\subsection{The one-fluid method}
\label{onefluidmodel}
The one-fluid equations can be derived by rewriting the fluid equations for the gas and the dust in the barycentric reference frame of the mixture \citep{2014MNRAS.440.2136L}. In doing so, we substitute out the individual velocities of the gas and dust phases in favour of the new barycentric velocity of the mixture,
\begin{equation}
{\bf v} =\frac{\rho_\textrm{g} {\bf v_{\rm g}}+\rho_\textrm{d}{\bf v_{\rm d}}}{\rho_\textrm{g}+\rho_\textrm{d}} ,
\end{equation}
and the differential velocity between the two phases, 
\begin{equation}
\Delta {\bf v} = {\bf v_{\rm d}} - {\bf v_{\rm g}} ,
\end{equation}
where $\rho$ is the density, ${\bf v}$ is the velocity, and the subscripts g and d identify gas and dust quantities, respectively. Similar to the velocities, we replace the gas and dust densities by the total density, $\rho = \rho_\textrm{d}+\rho_\textrm{g}$, and the dust fraction,
$\epsilon \equiv \rho_\textrm{d}/\rho$, such that
\begin{align}
\rho_{\rm g} &= (1-\epsilon) \rho  , \\
\rho_{\rm d} &= \epsilon \rho  .
\end{align}
The equations describing the evolution of a dust-gas mixture can be therefore written in the form \citep{2014MNRAS.440.2136L}
\begin{align}
\label{cont1flu}
& \frac{{\rm d} \rho}{{\rm d} t} = - \rho(\nabla \cdot \textbf{v}) , \\
\label{epsilon1flu}
& \frac{{\rm d} \epsilon}{{\rm d} t} = - \frac{1}{\rho} \nabla \cdot \left[\epsilon (1-\epsilon) \rho \Delta \textbf{v} \right] , \\
\label{vel1flu}
& \frac{{\rm d} \textbf{v}}{{\rm d} t} = -\frac{\nabla P}{\rho} - \frac{1}{\rho} \nabla \cdot \left[\epsilon (1-\epsilon) \rho \Delta \textbf{v} \Delta \textbf{v} \right] + {\bf f} ,  \\
\label{deltavel1flu}
& \frac{{\rm d} \Delta \textbf{v}}{{\rm d} t} = -\frac{\Delta \textbf{v}}{t_{\rm s}} + \frac{\nabla P}{(1-\epsilon) \rho} -(\Delta \textbf{v} \cdot \nabla)\textbf{v} + \frac{1}{2} \nabla \cdot \left[ (2\epsilon -1) \Delta \textbf{v} \Delta \textbf{v} \right] , \\
\label{ener1flu}
& \frac{{\rm d} \tilde{u}}{{\rm d} t} = - \frac{P}{\rho} \nabla \cdot (\textbf{v}-\epsilon \Delta \textbf{v}) +\epsilon (1-\epsilon) \frac{\Delta \textbf{v}^2}{t_{\rm s}},
\end{align}
where $P$ is the gas pressure and ${\bf f}$ represents the external forces acting on both components (e.g. gravity). Moreover, for convenience we have parametrised the thermal energy as $\tilde{u} = u (1-\epsilon)$. The stopping time, $t_{\rm s}$, is given by
\begin{equation}
\label{stoppingtime}
t_{\rm s}=\frac{\epsilon (1-\epsilon) \rho}{K} ,
\end{equation}
where $K$ is the drag coefficient, which regulates the aerodynamical coupling between the two phases \citep{weidenschilling77a}. The equations of the mixture are closed by the equation of state, such as the adiabatic one, i.e.
\begin{equation}
P = (\gamma-1) \rho \, \tilde{u} .
\end{equation}
There are several advantages to using the one-fluid formulation over the two-fluid approach (see \citealt{2015MNRAS.451..813P}), particularly for small dust grains. 
For example, since the gas and dust are co-located in the one-fluid approach, it does not require (or can easily circumvent) the prohibitive temporal and spatial resolution requirements at high drag (needed in two-fluid simulations by the interpolation of fluid quantities between different phases, \citealt{2012MNRAS.420.2345L}). Furthermore, the one-fluid method prevents artificial trapping of dust beneath the resolution length of the gas. Finally, the one-fluid formalism naturally generalises to account for multiple dust species coupled to the same gas phase \citep{2014MNRAS.444.1940L,hutchison18}.

\subsection*{Terminal Velocity Approximation}
The fluid equations in the one-fluid formalism can be simplified when the stopping time is small compared to the typical hydrodynamic timescale, i.e. the time required for a sound wave to propagate over a characteristic distance. In the context of SPH, we can write this condition as, $t_{\rm s} < h/c_{\rm s}$, where $h$ is the local smoothing length of the particles. In this regime, usually referred to as \emph{terminal velocity regime} \citep{2005ApJ...620..459Y},  the relative velocity between the two phases rapidly reach a terminal velocity due to the balancing of the drag and pressure forces. As a consequence, the time dependence of the differential velocity between the gas and the dust can be ignored,
\begin{equation}
\label{termvelapp}
\Delta {\bf v} = t_{\rm s} \frac{\nabla P}{\rho_{\rm g}} = \frac{t_{\rm s}}{(1-\epsilon)} \frac{\nabla P}{\rho}.
\end{equation}
Neglecting terms of second order in $t_{\rm s}$, Eqs.~\eqref{cont1flu}-\eqref{ener1flu} reduce to
\begin{align}
\label{eq:continuity}
& \frac{{\rm d} \rho}{{\rm d} t} = - \rho(\nabla \cdot \textbf{v}) , \\
\label{epsilonapp}
& \frac{{\rm d} \epsilon}{{\rm d} t} = - \frac{1}{\rho} \nabla \cdot \left(\epsilon t_{\rm s} \nabla P \right) ,\\
\label{}
& \frac{{\rm d} \textbf{v}}{{\rm d} t} = -\frac{\nabla P}{\rho} + {\bf f} ,  \\
\label{}
& \frac{{\rm d} \tilde{u}}{{\rm d} t} = - \frac{P}{\rho} ( \nabla \cdot \textbf{v}) .
\end{align}
Apart from the additional evolution equation for the dust fraction, the equations in the terminal velocity approximation bear striking resemblance to the usual hydrodynamic equations for the gas without the dust.
Therefore, the SPH discretisation of the continuity and momentum equations are identical to that of a regular gas-only simulation while the dust fraction and the thermal energy are discretised directly as shown in Eq.~43 in \cite{2015MNRAS.451..813P} and Eq.~55 in \cite{hutchison18}.

\subsection{Timestepping}
The addition of the diffusion equation for the dust fraction (Eq.~\ref{epsilonapp}) leads to an additional constraint on the timestep. Assuming a constant density and an isothermal equation of state, $P=c_{\rm s}^2 (1-\epsilon)\rho$, Eq.~\eqref{epsilonapp} can be rewritten as
\begin{equation}
\label{diffeqepsilon}
\frac{{\rm d} \epsilon}{{\rm d} t} = \nabla \cdot \left(\eta_{\epsilon} \, \nabla \epsilon \right),
\end{equation}
where $\eta_\epsilon \equiv \epsilon t_{\rm s} c_{\rm s}^2$ is the diffusion coefficient. A new constraint on the timestep is needed when the diffusion coefficient is larger. Indeed, \cite{2015MNRAS.451..813P} provide a stability criterion of the form
\begin{equation}
\Delta t < \Delta t_{\epsilon} \equiv C_0 \frac{h^2}{\eta_\epsilon} = C_0 \frac{h^2}{\epsilon c_{\rm s}^2 \, t_{\rm s}},
\label{eq:timestepconst}
\end{equation}
which implies that the timestep needs to be constrained when the stopping time is long -- the opposite of the two-fluid case where the timestep is constrained for short stopping times. It is worth remarking that the terminal velocity approximation is only strictly valid when the stopping time is less than the computational timestep. 
Actually, a more general timestep condition can be derived, taking into account possible gradients in $\epsilon$ (see Appendix~\ref{timestepgradeps}). This is given as:
\begin{equation}
\Delta t < C_0 \frac{h^2}{\epsilon c_{\rm s}^2 t_{\rm s}} \frac{2a}{a^2 + b^2} ,
\label{eq:timestepconst2}
\end{equation}
where $a=(1-h^2\nabla^2\epsilon/\epsilon)$ and $b=2h|\nabla\epsilon|/\epsilon$. It can be easily seen that Eq. (\ref{eq:timestepconst2}) reduces to Eq. (\ref{eq:timestepconst}) for constant $\epsilon$. This condition is safer than Eq. (\ref{eq:timestepconst}) in regions of strong gradients of $\epsilon$, but it is more difficult to implement (since it requires an additional loop over the particles to obtain the gradient of $\epsilon$) and can lead to severe timestep restrictions in certain practical applications (see Sect.~\ref{limstoppingtime}). As a result, we default back to Eq.~(\ref{eq:timestepconst}) for our timestep control in this work.

\section{Enforcing positivity of the dust fraction}
\label{sectrhoeps}
The one-fluid approach does not put any constraint on the positivity of the dust fraction. This problem can arise in regions where, for example, particles containing a finite amount of dust are adjacent to pure gas particles (i.e. $\epsilon = 0$). As the particles evolve in time, the infinite gradient in $\epsilon$ created at this interface leads the pure gas particles to develop a negative dust fraction. We can avert this problem by parameterising and evolving the dust fraction using a new variable, $s=\sqrt{\epsilon \rho}$. The positivity of the physical variable $\epsilon$ is now guaranteed since
\begin{equation}
\epsilon = s^2/\rho .
\end{equation}
The corresponding diffusion equation for the new variable $s$ is 
\begin{align}
\label{evolutionofs}
\frac{{\rm d} s}{{\rm d} t} &= -\frac{1}{2s} \nabla \cdot \left( \frac{s^2}{\rho} t_{\rm s} \nabla P \right) - \frac{s}{2} \nabla \cdot {\bf v} \nonumber \\
&=-\frac{1}{2} \nabla \cdot \left( \frac{s}{\rho} t_{\rm s} \nabla P \right) -\frac{t_{\rm s}}{2\rho} \nabla P \cdot \nabla s - \frac{s}{2} \nabla \cdot {\bf v} .
\end{align}
We note that the first term on the right hand side of Eq.~\ref{evolutionofs} is written so as to prevent an infinite gradient in $\epsilon$ when $s \rightarrow 0$ (i.e. $\epsilon \rightarrow 0$).
The usual method for discretising Eq.~\ref{eq:continuity},
\begin{equation}
	\label{}
	\rho_{a} = \sum_b m_b W_{ab} (h_a),
   \label{eq:rhoasum}
\end{equation}
trivially conserves the total mass of the mixture, but does nothing to conserve the mass of each of the components. Formally, mass conservation of the dust and gas also holds as long as the energy equation is modified appropriately \cite{2017arXiv170203930P}, i.e. such that
\begin{equation}
	\label{}
    \frac{\mathrm{d} E}{\mathrm{d} t} = \frac{\mathrm{d}}{\mathrm{d} t} \sum_a m_a \left[ \frac{1}{2} \mathbf{v}^2_a + (1-\epsilon_a) u_a \right] = 0,
\end{equation}
which, in terms of the new variable $s$, requires that
\begin{align}
	\label{}
    & \sum_a m_a \left[ \mathbf{v}_a \frac{\mathrm{d} \mathbf{v}_a}{\mathrm{d} t} + \rho^{\rm g}_a \frac{\mathrm{d} u_a}{\mathrm{d} t} - u_a \left( \frac{2 s_a}{\rho_a} \frac{\mathrm{d} s_a}{\mathrm{d} t} - \frac{s^2_a}{\rho^2_a} \frac{\mathrm{d} \rho_a}{\mathrm{d} t}\right) \right] = 0 .
\end{align}
The SPH discretisation for the evolution of $s$ is shown in Eq.~280 of \cite{2017arXiv170203930P}. Although the formulation prevents $\epsilon$ from going negative, it does not guarantee that the dust fraction will remain smaller than unity. Numerical artefacts can appear in regions where the gradient of the dust fraction is steep, resulting in a spontaneous increase in dust mass. These artefacts are most severe when $\epsilon \rightarrow 0$ or $\epsilon \rightarrow 1$ and, at least in some instances, quickly drive the dust fraction to values larger than unity.
\begin{figure*}
\centering
\includegraphics[width=\textwidth,trim={0.5cm 5.1cm 2.cm 0cm},clip]{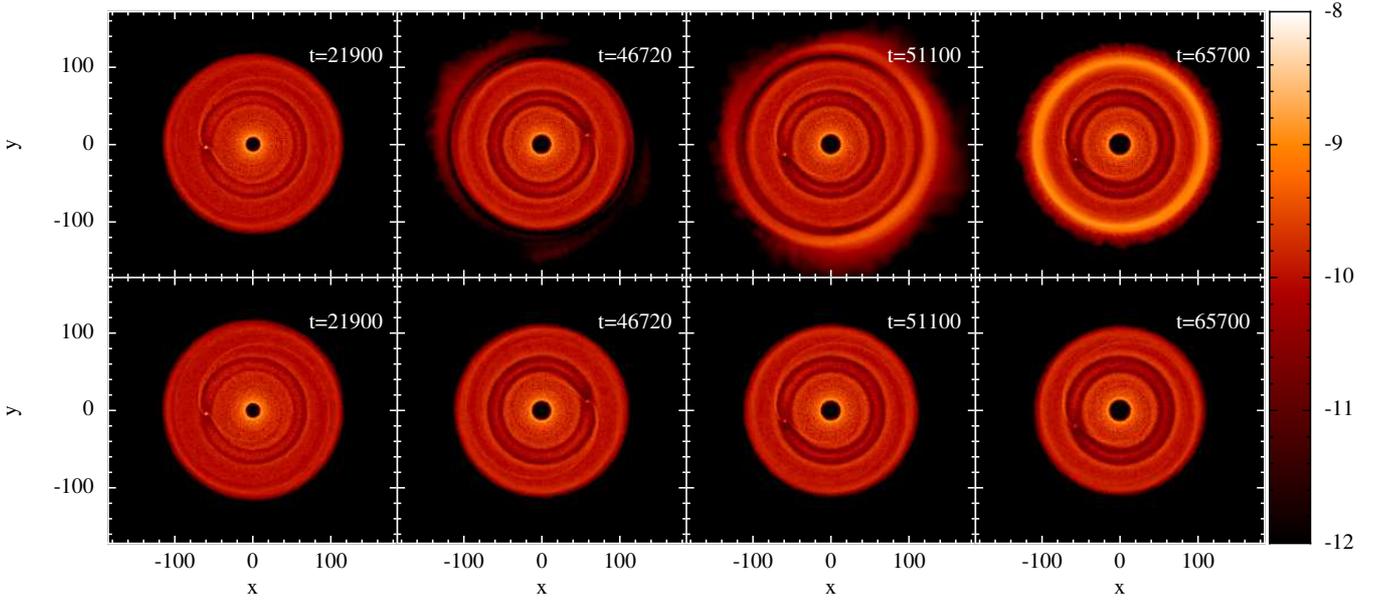}
\caption{Rendered images of dust surface density (in code units) at different times during a 3D SPH simulation of a dusty protostellar disc with a radial extent of $R \in [1,120]\;$ au and an embedded planet of mass 0.5 $\rm{M_J}$ at 60 au. The dust surface density profile follows a power law with index $p=0.5$. We used both the implementations described in  Sect.~\ref{sectrhoeps} (upper panels) and Sect.~\ref{sectepseps-1} (lower panels). The gas disc density structure (not shown) is spatially larger than the dusty disc, producing a region in the outer disc with a strong gradient in the dust diffusivity. The evolution of the dust dynamics in these regions is better handled with the new implementation. In particular, the spurious dust rings, that appear at late times with the old formulation and that signal that dust mass is not well conserved, disappear with the new formulation. The temperature profile drops as a power law with $q=-0.7$ and the disk aspect ratio is $H(R_0)/R_0=0.025$, at $R_0=1\,$au. The simulation describes the evolution of a $0.1$ millimeter grain population.}
\label{fig_sim}
\end{figure*}

\section{A new implementation}
\label{sectepseps-1}
In this section we propose a new parametrization of the dust fraction similar to that used by \cite{2015MNRAS.451..813P}, but that enforces the constraint $0 < \epsilon<1$ by mapping the dust fraction to a function whose co-domain is only defined from $[0,1]$, thereby preventing $\epsilon$ from becoming unphysical.
A promising parametrization that meets the above criterion is given by
\begin{equation}
\epsilon=\frac{s^2}{1+s^2} \quad \text{such that} \quad s=\sqrt{\frac{ \epsilon}{1-\epsilon}}.
\end{equation}
In this new formulation the variable $s$ is then related simply to the ratio of dust to gas densities, $s=\sqrt{\rho_{\rm d}/\rho_{\rm g}}$. We calculate the time derivative as
\begin{equation}
\label{evols}
\frac{{\rm d} s}{{\rm d} t}= \frac{1}{2s(1-\epsilon)^2} \frac{{\rm d} \epsilon}{{\rm d} t} .
\end{equation}
Substituting Eq.~\eqref{epsilon1flu} and manipulating the  term on the right hand side of Eq.~\eqref{evols}, we obtain
\begin{align}
\label{dsoverdt}
\frac{{\rm d} s}{{\rm d} t} = -\frac{1}{2\rho(1-\epsilon)^2} \biggl \lbrace \nabla\cdot\bigl[s(1-\epsilon)  t_{\rm s} \nabla P\bigr] + (1-\epsilon)t_{\rm s} \nabla P \cdot \nabla s \biggr \rbrace .
\end{align}
The SPH discretisation is implemented in the form
\begin{align}
\frac{{\rm d} s_a}{{\rm d} t} = - \frac{1}{2\rho_a \left(1-\epsilon_a \right)^2} \sum_b \biggl [ \frac{m_b s_b}{\rho_b}\left(D_a +D_b\right)(P_a-P_b)\frac{\overline{F}_{ab}}{|r_{ab}|} \biggr ],
\end{align}
where $D_a \equiv t_{{\rm s},a}\left(1-\epsilon_a \right)$.
Like the previous implementation (Sect.~\ref{sectrhoeps}), our new expressions conserve linear and angular momentum, energy, and mass --- at least up to the accuracy of the timestepping algorithm. Although it is true that the \emph{total} mass is trivially conserved by virtue of Eq.~\eqref{eq:rhoasum}, this attribute is not bequeathed to the individual phases due to their dependence on $\epsilon$, an evolved quantity. This contingency on the time-evolution accuracy of $\epsilon$ plays an important role in the discussion that follows.

We implemented the above formalism into the SPH code \textsc{Phantom} \citep{LP10,2017arXiv170203930P} and tested it using \textsc{Phantom}'s standard nightly test suite (described in Sect. 5.1 of \citealt{2017arXiv170203930P}), which includes (among others) the \textsc{dustywave}, \textsc{dustyshock}, and \textsc{dustydiffuse} tests described in \citet{2015MNRAS.451..813P}. The new implementation not only passed within the `acceptable' tolerances set for each test, it outperformed the existing algorithm. As a specific example, when compared with the previous method, the `derivatives test' (Sect. 5.1.1 of \citealt{2017arXiv170203930P}) showed an improvement in the accuracy of the time derivative of $\epsilon$ by a factor of five while the total energy conservation improved by a factor of $\sim400$.

Next we looked at some typical configurations involving the interaction of an embedded protoplanet with its parent disc.
Comparing the two parametrizations discussed in this paper, Fig.~\ref{fig_sim} follows the evolution of the dust surface density (initial power law profile with index $p = -0.5$) in a 3D simulation of a dusty protostellar disc with a radial extent of $R \in [1,120]\;$au and an embedded planet of mass 0.5 $\rm{M_J}$ located at a distance of 60 au from the central star. The temperature profile drops as a power law with $q=-0.7$ and the disc aspect ratio is $H(R_0)/R_0=0.025$, at $R_0=1\,$au. We embed the planet in order to further investigate diffusivity gradients that arise due to planet-disc interactions. The planet also alters the relative dust fractions in the inner and outer parts of the disc with time. The simulation describes the evolution of a $0.1$ millimeter grain population. Particles with a non-negligible dust fraction exhibit Stokes numbers in the range $[0.02,0.2]$, which safely correspond to stopping times below $h/c_{\rm s}$.

As time progresses, the viscous and pressure forces in the disc cause the gas to expand radially outward, creating a strong gradient in the dust fraction (and hence diffusivity) at the edge of the dusty disc. 
Fig.~\ref{fig_sim} shows that the numerical artefacts that occurred with the old implementation are removed with the new parametrization. This improved accuracy is thanks to the more accurate time-evolution of $\epsilon$ in regions with steep gradients in the dust diffusivity (i.e., at the outer edge of the dusty disc).
\begin{figure}
\includegraphics[width=0.5\textwidth]{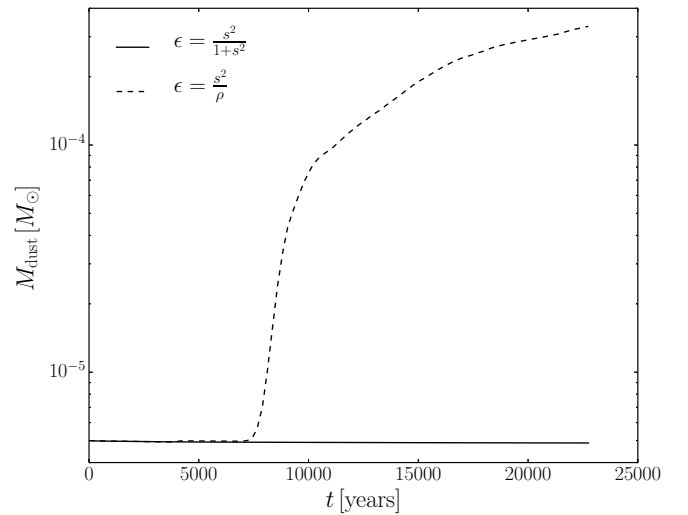}
\caption{Time evolution of the total dust mass for the parametrization $\epsilon=s^2/\rho$ described in Sect.~\ref{sectrhoeps} (dashed line) and $\epsilon=s^2/(1+s^2)$ in Sect.~\ref{sectepseps-1} (solid line). Importantly, the dust mass does not increase over time with the new parametrization.}
\label{fig:dust-cons}
\end{figure}

Again comparing the two implementations, Fig.~\ref{fig:dust-cons} shows the time evolution of the total dust mass, i.e $\epsilon=s^2/\rho$ (dashed line) and $\epsilon=s^2/(1+s^2)$ (solid line). While with the old implementation the dust mass increases in time (starting from a value of $5\cdot 10^{-6}\, M_{\odot}$ and reaching $3 \cdot 10^{-4}\, M_{\odot}$, after $\sim 2\cdot 10^{4}$ years), the new implementation better computes the evolution of the dust density, avoiding most of the numerical artefacts occurring at the edge of the dusty disc due to the strong gradients in the dust fraction.
Moreover, our tests show that the computation of the dust fraction and the thermal energy in our new implementation is faster than the $\sqrt{\epsilon \rho}$ parametrization described in Sect.~\ref{sectrhoeps}. 

\section{Limiting the stopping time}
\label{limstoppingtime}
As mentioned earlier, despite the conservation ensured by the spatial discretisation of the fluid equations, non-conservation may still arise due to timestepping errors. Non-conservation of gas/dust mass are particularly vulnerable in regions of small $\epsilon$ where the dust fraction tends to relax the timestep (see Eq.~\ref{eq:timestepconst}). However, since these regions are usually occupied by dust grains with large stopping times, they are the very regions that need a small timestep in order to be accurate. This breakdown of our timestep criterion is most likely due to the violation of the assumptions used to derive Eq.~(\ref{eq:timestepconst}), and in particular to the fact that it was derived neglecting gradents in the dust fraction, as discussed already in Sect.~\ref{sectrhoeps} above. In theory, we should be able to reduce our timestep (by adopting the full timestep condition, Eq.~\ref{eq:timestepconst2}, or by reducing $C_0$) to maintain mass conservation. We have verified that maintaining a `sufficiently small' timestep for these problematic particles preserves mass conservation for the system, but at the cost of impossibly slow simulations when, e.g., very small amounts of dust get flung out and stranded in the low-density outer disc. Therefore, in practice we seek a more viable option that can circumvent these problem particles while still conserving gas/dust mass for the system.
It is rather vexing that such violations most likely occur in `peripherial' particles that often have little influence on the simulation at large. From experience, numerical artefacts are mostly likely to occur in the upper/outer regions of discs with high aspect ratio, $H/R$, and low (in absolute value) radial power-law index for the temperature, $q$. The dust diffusion, i.e. $\epsilon t_{\rm s} \nabla P$, in these regions is strong due to the steep gradients in the pressure and for particles with large stopping time. 

To prevent the numerical inaccuracies we see when such strong gradients are present in the disc for particles with large stopping time, we propose moderating the rapid dust diffusion for problematic particles by enforcing the following limit on the stopping time
\begin{equation}
\label{eq:ts_limited}
\tilde{t}_{\rm s} = {\rm min}\left(t_{\rm s},h/c_{\rm s} \right) ,
\end{equation}
that results in limiting the flux of the mass embodied in large particles.
Limiting the flux of dust mass through the stopping time (as opposed to the pressure gradient or the diffusion coefficient as a whole) has the advantage that it is localised strictly to particles that violate the terminal velocity approximation and requires no prior knowledge about the dynamical state of the system. 

\begin{figure*}
\centering
\includegraphics[width=\textwidth,trim={0.5cm 4cm 2cm 0cm},clip]{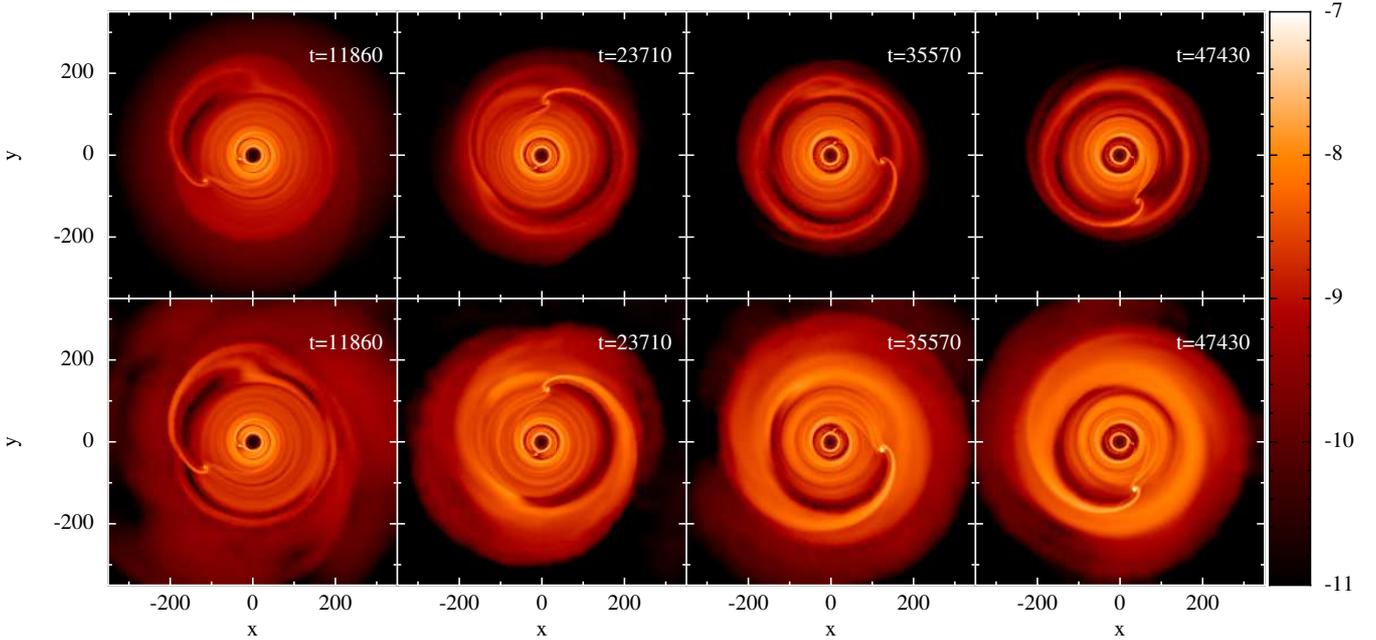}
\caption{Rendered images of dust surface density (in code units) at different times using our new $\sqrt{\rho_{\rm d}/\rho_{\rm g}}$ parametrization, including the limit on the stopping time (top panels) and not (bottom panels).
To test the limits of our algorithm, we alter the disc model so that a large fraction of the dust grains in the outer disc have a stopping time larger than $h/c_{\rm s}$. We further exacerbate the conditions by placing a massive protoplanet near the outer disc edge to stir up the dust in low density regions. Limiting the stopping time allows mass conservation to hold, even in these extreme conditions. The protoplanetary disc used in this simulation has a radial extent of $r \in [25,200]\,$au, with an aspect ratio, $H(R_0)/R_0=0.09$, at $R_0=25\,$au and a power law temperature profile with index $q=-0.5$. The gaseous disc mass is $0.034\,M_\odot$, with a dust-to-gas ratio of $0.007$. The initial gas and dust surface densities are given by a power law (index $p=-1$) with an exponential cut-off at $R_{\rm tap}=70$ au. We considered a dust grain size of $1\,{\rm mm}$. Two planets are embedded in the disc at $35\,$au and $140\,$au, of $4\,\rm{M_J}$ and $6\,\rm{M_J}$, respectively.
}
\label{fig_simlimiter}
\end{figure*}
Fig.~\ref{fig_simlimiter} compares the evolution of the dust surface density using our new dust implementation presented in Sect.~\ref{sectepseps-1} (lower panels) and the same implementation, but limiting the stopping time (upper panels). The protoplanetary disc used in these two simulations has a radial extent of $r \in [25,200]\,$au and it is thicker than the one used in Fig.~\ref{fig_sim}, with an aspect ratio, $H(R_0)/R_0=0.09$, at $R_0=25\,$au and a temperature profile index $q=-0.5$. The gaseous disc mass is $0.034\,M_\odot$, with a dust-to-gas ratio of $0.007$. The initial gas and dust surface densities are given by a power law (index $p=-1$) with an exponential cut-off at $R_{\rm tap}=70$ au. The dust grain size is $1\,{\rm mm}$. In this case, we include two planets at $35\,$au and $140\,$au, of $4\,\rm{M_J}$ and $6\,\rm{M_J}$, respectively. The outer planet is deliberately placed so as to fling dust into regions where we know the terminal velocity approximation has difficulty. Importantly, the flux limited simulations conserve the dust mass to machine precision while our other simulations do not. The spurious increase in dust mass in our unmodified simulation takes place in the outer disc where the gradients in the dust diffusivity are large.

It is important to note that by limiting the stopping time we are artificially modifying the Stokes number. Rewriting Eq.~\ref{eq:ts_limited} in terms of the Stokes number (for discs in vertical hydrostatic equilibrium, i.e. $H = c_{\rm s}/\Omega_{\rm k}$) yields
\begin{equation}
{\rm St} < h/H .
\end{equation}
Since in typical SPH simulations $h/H<1$, this new implementation affects the dust density evolution of large dust grains, even with moderately low $\mathrm{St}>h/H$. Since the radial dust velocity increases with $\mathrm{St}$ for $\mathrm{St}<1$ \citep{nakagawa86a}, limiting the stopping time leads to an underestimate of the radial flux of large grains towards disc regions of high pressure. Consequently, the new-found mass conservation afforded by limiting the flux is not an excuse to apply our method in every situation. In particular, care should be taken when simulating protoplanetary discs with high aspect ratio and low $q$, where it is more likely to find dust grains with both large and small Stokes number. For these discs, a correct physical description of the system may only be attainable with the full one-fluid approach \citep{2014MNRAS.440.2136L}, hybrid method combining the one- and two-fluid approaches or semi-analytical two-fluid methods \citep[e.g.][]{aguilarbate14}. 

In summary, limiting the stopping time conserves dust mass and prevents numerical artefacts from developing in particles in the outer disc where Stokes numbers are large  and dust mass is negligible. The evolution of the dust density in these situations can be considered reliable. However, when the stopping time of particles are being limited in the bulk of the disc, where mass fractions are still high, we recommend using a different approach.

\section{Conclusions}
We introduce a new algorithm to compute the dynamics of tightly-coupled dust grains in the context of the one fluid approach described in \citet{2014MNRAS.440.2136L}. Our algorithm avoids certain numerical artefacts that arise in the previous formalism \citep{2015MNRAS.451..813P}, rendering our method both faster and more accurate. We do this by
\begin{enumerate}[itemsep=1mm, topsep=5.pt, leftmargin=.25in]
\item [(i)] parameterising the dust fraction using the square root of the dust-to-gas ratio, which enforces $\epsilon \in [0,1]$;
\item [(ii)] limiting the stopping time below a value that ensures the validity of the equations of motion in the terminal velocity approximation, i.e. $t_{\rm s} < h/c_{\rm s}$.
\end{enumerate}
The latter leaves the numerically-stable, strongly-coupled dust grains untouched, while limiting the amount of dust that can be transferred between weakly-coupled particles that would otherwise violate the assumptions of the one-fluid diffusion approximation. When the flux in these weakly-coupled grains is not constrained, the dust mass can un-physically grow over long times in some regions of the disc, violating mass conservation. We find no adverse effects of limiting the flux of particles with low dust fraction, which are typically found in the upper/outer regions of the disc. However, we caution that the stopping time limiter needs to be used with care, since it can lead to an incorrect computation of the dust dynamics of large decoupled dust grains when the dust fraction is non-negligible. In these situations, we recommend switching to a two-fluid formalism \citep{2012MNRAS.420.2345L,aguilarbate14}. 

Finally, there are realistic scenarios in which a single grain size can be strongly-coupled in one region of the disc and weakly-coupled in another -- with a significant dust mass in each region. In this scenario, neither the one-fluid diffusion approximation or the two-fluid method would be adequate, but would require a hybrid scheme that marries the two approaches or, alternatively, the full one-fluid formalism that allows for a wider range in drag regimes \citep{2014MNRAS.440.2136L,2014MNRAS.440.2147L}. Alternatively, implicit or semi-analytic methods have been proposed to simulate tightly-coupled particles in multi-fluid simulations with strong drag regimes \citep{aguilarbate14,booth15,aguilarbate15}.

\section*{Acknowledgments}
We are grateful to an anonymous referee for constructive suggestions that improved our paper.
The authors would like to thank Richard Alexander and Chris Nixon for fruitful discussions. 
This project has received funding from the European Research Council (ERC) under the European Union's Horizon 2020 research and innovation program (grant agreement No 681601).
This research used the ALICE High Performance Computing Facility and the DiRAC Complexity system, funded by BIS National E-Infrastructure capital grant ST/K000373/1 and  STFC DiRAC Operations grant ST/K0003259/1. We used \textsc{SPLASH} \citep{splash}.




\bibliographystyle{mnras}
\bibliography{mybibfile} 




\appendix



\section{Time stepping with gradients of $\epsilon$}
\label{timestepgradeps}

For simplicity, we derive the time step condition with non-zero derivatives of the dust fraction in 1D first, from
\begin{equation}
\frac{\mathrm{d}\epsilon}{\mathrm{d} t} = \frac{\partial }{\partial x} \left( \epsilon t_{\rm s} c_{\rm s}^{2} \frac{\partial \epsilon }{\partial x}\right) .
\label{eq:diff_1D}
\end{equation}
For the first order backward Euler scheme, the linear expansion of Eq.~\ref{eq:diff_1D} for modes of the form $\epsilon^{n} = \epsilon_{0} + \delta \epsilon^{n} \mathrm{e}^{ikx}$ provides 
\begin{equation}
\delta \epsilon^{n + 1} = \delta \epsilon^{n}\left[  1 + \Delta t \left( t_{\rm s} c_{\rm s}^{2}\right) \left( -k^{2}\epsilon_{0} + \frac{\partial^{2} \epsilon_{0}}{\partial x^{2}} + 2i k \frac{\partial \epsilon_{0}}{\partial x }\right)   \right] .
\end{equation}
The numerical scheme requires $\left| \delta \epsilon^{n + 1} / \delta \epsilon^{n} \right| < 1$ for stability. With the usual substitution $k \to h^{-1}$, this condition gives
\begin{equation}
\left| 1 - q\left( a + ib \right)  \right|  < 1 ,
\end{equation}
where $q \equiv \Delta t /\left(  h^{2} / \eta_{\epsilon,0} \right) $, $a \equiv 1 - \frac{h^{2}}{\epsilon_{0}} \frac{\partial ^{2}\epsilon_{0}}{\partial x^{2}}  $ and $b \equiv 2 \frac{h}{\epsilon_{0}}\frac{\partial \epsilon_{0}}{\partial x}$. Hence,
\begin{equation}
\left( 1 - qa \right)^{2} + \left( qb\right)^{2} < 1.
\label{eq:ineq_inter}
\end{equation}
Expanding the left-hand side of Eq.~\ref{eq:ineq_inter} and dividing by $q>0$ provides finally
\begin{equation}
q < \frac{2a}{a^{2} + b^{2}}.
\label{eq:final_ineq}
\end{equation}
Putting a safety constant $C_{0}$ in front of the right-hand side of Eq.~\ref{eq:final_ineq} gives the generic form for the time step condition with gradients of $\epsilon$, which can be generalised in 3D accordingly.


\bsp	
\label{lastpage}
\end{document}